\documentclass[12pt]{iopart}
\usepackage{graphicx,cite}

\newcommand{\etc}{{\it etc.}}
\newcommand{\YBCO}{YBa$_2$Cu$_3$O$_{7-\delta}\ $}
\newcommand{\SmBCO}[1]{SmBa$_2$Cu$_3$O$_{#1}$}
\newcommand{\AS}{{\it Appl. Supercond.}\ }
\newcommand{\IEEESup}{{\it IEEE Trans. Appl. Supercond.}\ }
\begin{document}

\title[Growth-related profiles of remanent flux in bulk melt-textured 
YBaCuO crystal]{Growth-related profiles of remanent flux in bulk 
melt-textured YBaCuO crystal magnetized by pulsed fields}
\author{A~B~Surzhenko\dag\ddag, S~Schauroth\dag, D~Litzkendorf\dag, 
M~Zeisberger\dag, T~Habisreuther\dag\ and W~Gawalek\dag} 
\address{\dag\ Institut f\"ur Physikalische Hochtechnologie, Winzerlaer 
Str. 10, Jena D-07745, Germany} \address{\ddag\ Institute for Magnetism, 
Vernadsky Str. 36b, Kyiv 03142, Ukraine}

\begin{abstract}
We have studied the remanent magnetic flux distribution in bulk 
melt-textured  \YBCO (YBCO) crystals after their magnetization in 
quasi-static and pulsed magnetic fields up to $6\,T$. It has been shown 
that, provided that the magnetic pulse is sharp enough and its amplitude 
much exceeds the twice penetration magnetic field, the pulse magnetization 
technique becomes extremely sensitive to the sample inhomogeneities. Using 
this method with appropriate parameters of the magnetic pulse, we have 
particularly demonstrated that the growth of YBCO crystals in the growth 
sectors responds for a macroscopic arrangement of weaks links. 
\end{abstract}



\date{\today}

\section{Introduction}
The recent progress in melt processing enables to fabricate large 
RE--Ba--Cu--O (RE is the rare-earth element) high-temperature 
superconductors (HTS) that look very promising for manifold applications, 
viz., for HTS permanent magnets, magnetic bearings and flywheels, 
electromotors and generators, current limiters, \etc \cite{Hab98} Thus, it 
is vitally important to find fast, sensitive and non-destructive method(s) 
for the quality tests of the HTS material. 

Now the remanent magnetic flux profiles \cite{1,3} of the field-cooled 
(FC) or zero-field-cooled (ZFC) samples are prevalent for these tasks. 
Compared to permanent magnets \cite{3}, strong magnetic fields of 
superconducting coils result in an increased sensitivity to the sample 
inhomogeneities. However, this seems to be essentially improved with the 
pulsed field magnetization (PFM). Till nowadays this technique was mainly 
considered as a convenient method to magnetize the HTS permanent magnets 
if these are preinstalled in various equipment such as, for example, 
electromotors \cite{Itoh0}. The main problem, which was encountered by 
various authors \cite{Itoh1,6,7,Mizutani,Sandler}, consisted in the 
trapped flux losses related to the sample heating by viscous forces that 
exert on moving flux lines \cite{Tsuchimoto}. Thus, the main pains were 
taken to reduce this destructive effect. In particular, various parameters 
of magnetic pulses (i.e. the shapes, durations and amplitudes) 
\cite{Itoh1,6,7} and/or their sequences \cite{Mizutani,Sandler} were 
tested for these purposes. However, being harmful for practical 
magnetization of HTS bulks, this phenomenon promises a certain profit for 
their characterization. In fact, the local areas containing weak links 
have been shown to be more heated than those where weak links are absent 
\cite{Itoh1}. If the magnetic pulse is short enough, the heat transfer 
during the pulse is nearly negligible. In this case, the temperature 
distribution just after the pulse has to correlate with the material 
quality and yet more enlarge the contrast between the magnetic flux 
trapped by areas with and without weak links. Thus, the recipe of how to 
increase this contrast seems simple: one has to accelerate the magnetic 
flux motion and to shorten the pulse duration. But despite of evident 
benefits for the HTS sample characterization, the PFM method did not gain 
a respective reputation so far. Moreover, no detailed studies: Which 
parameters of the magnetic pulse are enough to provide the required 
sensibility? were performed. 

Meantime, a macroscopic arrangement of structural imperfections (e.g., 
subgrains and 211-inclusions) which is related to the growth of the HTS 
crystals was recently inquired by polarized light methods \cite{Diko}. Due 
to extremely short coherence length $\xi$ which is typical for HTSs, this 
arrangement may certainly be expected to result in the growth-related 
magnetic properties. Except for an apparent importance of this feature 
itself, its detection by the PFM technique could serve as the best 
criterion whether a necessary sensitivity is already attained. To our 
knowledge, no growth-related profiles of the remanent magnetic flux were 
reported hitherto. That is what we are going to demonstrate in the current 
paper. 

\section{Experimental details} 
\begin{figure}[!tbp]\begin{center}
\includegraphics[height=\textwidth,angle=-90]{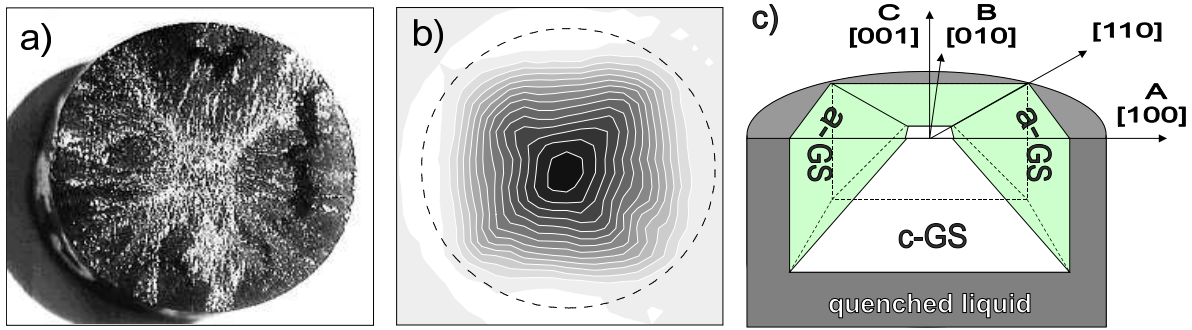}
\end{center}
\caption{The bulk YBCO sample that was studied in the current work. The 
$90^\circ$-cross on (a) the optical image and nearly regular tetragonal 
shape of isolines on (b) the remanent magnetic flux map (FC magnetization 
at T=77\,K and $\mu_0H$=1\,T, peak value $B_{max}$=0.63\,T) testify to the 
homogeneity of the crystal structure which sketch is illustrated in 
figure\,(c).} \label{sample}
\end{figure}

To avoid unwanted consequences that could 
be induced by cracks and other macroscopic imperfections, we have 
knowingly selected rather homogeneous \YBCO\ (YBCO) sample 
(\Fref{sample}). This was sinthesized from the mixture of commercially 
purchased powders (Solvay GmbH, Germany) with 1\,wt.\% CeO$_2$ by the 
melt-processing technique \cite{technology} and post-annealed in oxygen 
atmosphere. The bulk thus prepared was a cylinder with a diameter of 
30\,mm and a thickness of 18\,mm. To reduce a plausible scatter of flux 
distribution caused by a roughness of the top surface, the sample was 
polished. 

The optical image of the polished surface (see \Fref{sample}(a)) clearly 
exhibited a trace in its geometric centre, i.e. under the point where the 
\SmBCO{X} seed was placed, and the X-like cross separating the crystal 
onto four 90-degree segments. These features correspond to five growth 
sectors (GSs), i.e. areas grown on different habit planes: the c-GS with 
the habit (001) perpendicular to the c-axis, [001], and four a-GSs with 
the (100), (010), ($\bar{1}$00) and (0$\bar{1}$0) habit planes (see the 
schematic illustration in \Fref{sample}(c)) \cite{Diko}. In other words, 
both the central trace and the cross display the GS boundaries (GSBs), 
viz., a-c-GSBs and a-a-GSBs, respectively, which are pathways of crystal 
edges between two neighboring habits during the growth stage. Although a 
regular shape of the remanent flux map presented in \Fref{sample}(b) well 
confirmed the sample homogeneity, we shall hereafter report in what 
devastating contrast may minor inhomogeneities result under the PFM. 

In order to magnetize the sample we used either quasi-static fields with a 
sweep rate $V=\mu_0dH/dt\simeq 0.01\,T/s$ or the sin-wave-shaped magnetic 
pulses ($V\leq 6000\,T/s$). In all cases, the sample was magnetized in the 
vertical direction, i.e. along the c-axis, [001]. The static magnetization 
(both in the FC and ZFC modes) was provided by the iron-yoke magnet 
($\mu_0H\leq1\,T$) and the superconducting solenoid ($\mu_0H\leq8\,T$). 
Magnetic pulses were supplied by discharging of capacitors (with a total 
capacity of 5\,mF) charged up to 2\,kV through a circuit consisting of a 
copper coil and a thyristor which flashed only the first half 
($0\leq\omega t\leq\pi$) of a sin-wave $H=H^*\sin(\omega t)$. This form, 
amplitude ($\mu _0H^*\leq 6\,T$) and duration ($t_{max}=\pi/\omega=3\,ms$) 
were wittingly selected so to attain better sensitivity to the sample 
inhomogeneities. In our view, the sweep rates
$V=\pi\mu _0H^*\cos(\omega t)/t_{max}$, 
which are especially high both in the ascending ($t\simeq 0$) and, in 
contrast to the exponentially decaying pulses \cite{Itoh1,7,6,Mizutani}, 
descending ($t\simeq t_{max}$) branches, have to result in the heat 
generation which is almost twice larger than this registered in 
\cite{Itoh1,7,6,Mizutani}. In fact, numerical calculations 
\cite{Tsuchimoto} confirm that, owing to lower sweep rates, the descending 
brunch of the exponentially decaying pulses gives a minor contribution 
($\simeq 13\%$, i.e. $0.6\,K$ at $\mu_0H^*=4\,T$) to the total temperature 
change ($\Delta T=4.55\,K$ at $\mu_0H^*=4\,T$). 

However by simple heating of the sample, one could scarcely improve a 
sensibility. The heat also has to retain in the areas wherein this arises, 
i.e. in the areas with an excess of weak links \cite{Itoh1}. Short length 
of magnetic pulses promises to resolve this problem in the most important 
stage, i.e. during a pulse. Nothing can stop the heat exchange within the 
HTS sample afterwards, but the ``after-the-pulse'' smearing of the heat 
distribution is expected to reduce the remanent flux contrast less 
considerably. Besides, to minimize the ``after-the-pulse'' effects, for a 
whole experimental cycle (magnetizing $\rightarrow$ dwelling time of 
300\,s $\rightarrow$ measurements) the sample was immersed in a liquid 
nitrogen which, to our opinion, could provide a good thermal conductivity 
with, at least, the sample surfaces giving a main contribution to the 
measured flux density $B$.

The value of $B$ was mapped by the Hall sensor (Arepoc Ltd., Slovakia) 
which moved horizontally, 0.8\,mm above the surface, stepwise with a pitch 
of 1\,mm.  

\section{Results and discussion} 
\begin{figure}[!tbp] \begin{center}
\includegraphics[width=5cm,angle=-90]{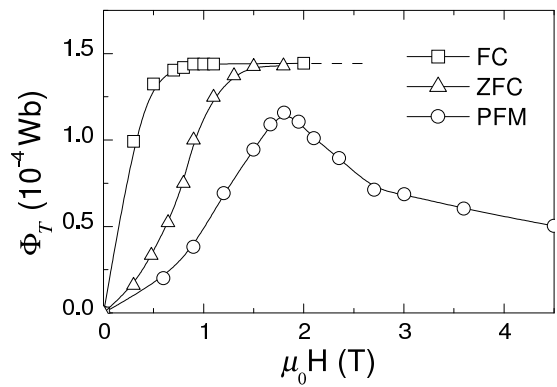} \end{center}
\caption{The applied magnetic field dependence of the total flux $\Phi_T$
trapped by the sample in the FC, ZFC and PFM modes (T=77\,K). In the last 
case, $\Phi_T$ is plotted vs the pulse amplitudes $\mu_0H^*$.}
\label{fig2}
\end{figure}

Before an immediate presentation of the 
remanent flux profiles, we have, at first, to note the following. The 
common-accepted value which characterizes the influence of the magnetic 
field on the sample ability to trap magnetic flux is the total  flux 
$\Phi_T$. To calculate $\Phi_T$ one usually integrates the measured flux 
density $B$ over an entire surface $S$ where $\Phi$ is positive. Although 
the averaged value $\Phi_T=\bar{B}S$ describes the sample only as a whole, 
it would nevertheless be useful to compare the data obtained under our 
pulse conditions with those reported in references 
\cite{Itoh1,6,7,Mizutani}. \Fref{fig2} presents such dependencies for the 
sample magnetized in the FC, ZFC and PFM modes. Obviously, the optimal 
field $\mu_0H_{opt}=1.9\,T$, where $\Phi_T$ in the PFM mode exhibits a 
maximum, corresponds to a full magnetization, i.e. the double penetration 
field $H_p=H_{opt}/2$. 

In general, these curves are similar to those obtained elsewhere 
\cite{Itoh1,6,7}, but a certain distinction of our PFM data does seem 
worthy of discussion. This feature, viz., much lower ratio 
$R(H>H_{opt})=\Phi^{PFM}_T/\Phi^{ZFC}_T$ of the trapped flux to its 
``isothermal'' ($\Delta T=0$) value (i.e. this obtained under the ZFC 
magnetization), is readily explained by more rapid motion of flux lines 
and, due to their interaction with viscous forces, more energy dissipated 
in the sample during a pulse. For example, $R\approx 0.65$ which was 
reported at $\mu_0H=4\,T$ in Ref.\,\cite{6} nearly twice exceeds similar 
value in Fig.\ref{fig2}, $R=0.38$. Assuming that the total flux $\Phi_T$ 
is proportional to the critical current density $j_c$ which linearly 
decreases with the temperature and vanishes at $T_c\approx91\,K$ 
\cite{tobepublished}, one can roughly estimate the temperature rise as 
$\Delta T=(1-R)\times(T_c-77\,K)$, i.e. $\Delta T=8.7\,K$ and $\Delta 
T=4.9\,K$ for $R=0.38$ and $0.65$, respectively. This naive approach 
appears, nevertheless, rather accurate: the latter value is in an 
agreement with $\Delta T\approx 4.6\,K$ calculated by authors \cite{6} 
from direct measurements of flux line velocities as well as with $\Delta 
T=4.55\,K$ given by numerical methods \cite{Tsuchimoto}. 
 
\begin{figure}[!tbp] \begin{center}
\includegraphics[angle=-90,width=0.8\textwidth]{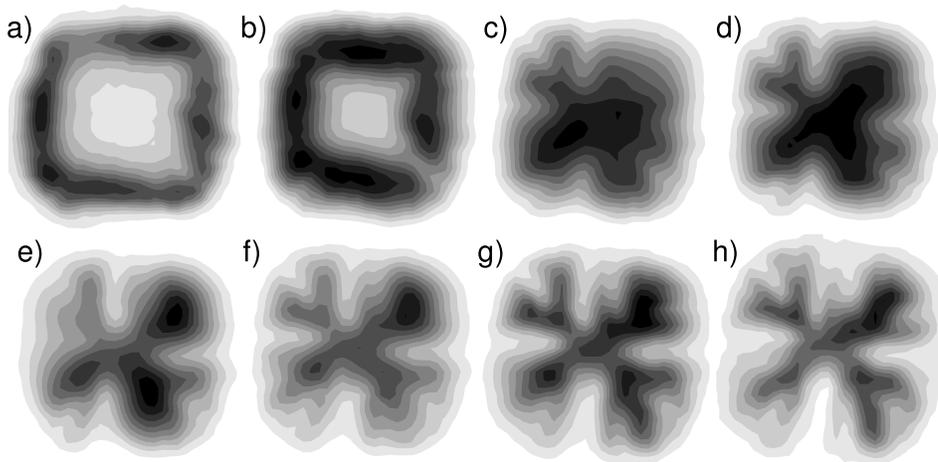} \end{center}
\caption{The trapped magnetic flux distributions for various
amplitudes of the pulsed field. The figures (a)-(h) correspond to
T=77\,K and $\mu_0H$=0.90, 1.50, 1.65, 1.95, 2.1, 2.4, 3.0 and 4.0\,T,
respectively. The different darknesses mark the 10\%-levels of the
flux density (zero levels are white, maximum values, $B_{max}$, are 
black).}
\label{fig3}
\end{figure}
However, large temperature rise $\Delta T\simeq 8.7\,K$ does not yet 
guarantee whether the generated heat distribution reflects the HTS quality 
or not. The answer may be obtained from \Fref{fig3} which shows the 
changes of the remanent flux distribution as $\mu_0H$ grows. To make this 
evolution more evident, we have previously normalized the flux densities 
$B$ for each measurement cycle to their peak value $B_{max}$. Owing to 
this trick, one can easily observe how the ring-like distributions 
gradually transform into the ``cross-wheels'' which ``rungs'' have the 
form of a dovetail. If the flux maps (a) and (b) are quite typical for the 
ZFC procedure at $H<H_{opt}$, the high-field data, $H>H_{opt}$, seem much 
more interesting. At first, so large gradients $dB/dx$ and $dB/dy$, where 
$x$ and $y$ are the space coordinates, as those shown in Figures 
\ref{fig3}(e)-(h) do confirm the above suggestion that the heat transfer 
within the HTS bulk is relatively small. But the main result is still the 
following: in view of the mentioned above scenario \cite{Itoh1}, the 
obtained flux profiles convincingly attest that weak links are mostly 
distributed inside the a-GSs, while a-a-GSBs are nearly free of them. 
Meanwhile, this arrangement is hardly noticeable under the slow 
magnetization. Upon a closer view, \Fref{sample}(b) displays this 
difference only through minor twists of isolines (confer, for example, 
their pincushion distortion along the top-right a-a-GSB with the maximum 
in Figures\,\ref{fig3}\,(e)-(h)). 

\begin{figure}[!tbp] \begin{center}
\includegraphics[angle=-90,width=0.36\textwidth]{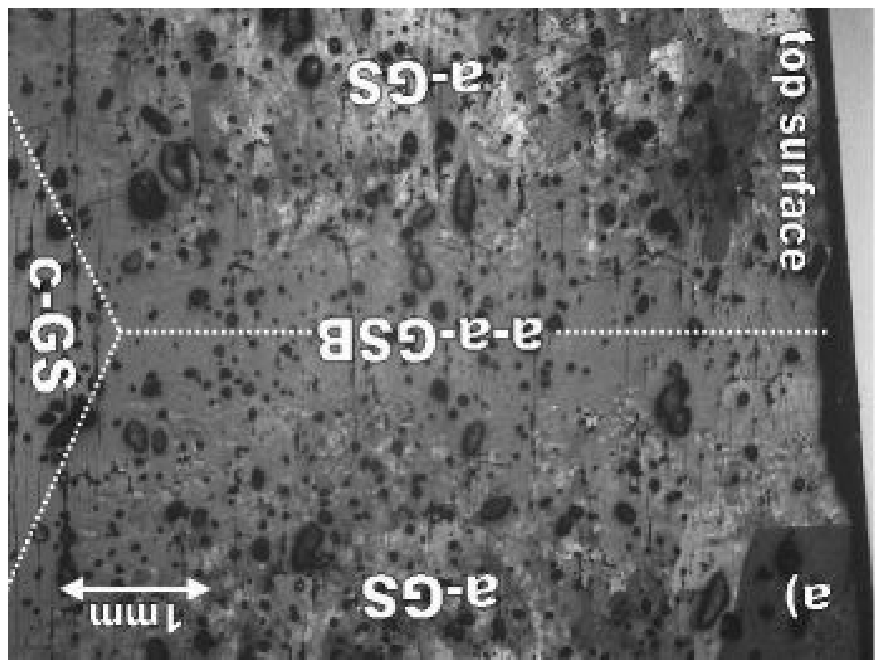} 
\hskip0.01\textwidth
\includegraphics[angle=-90,width=0.23\textwidth]{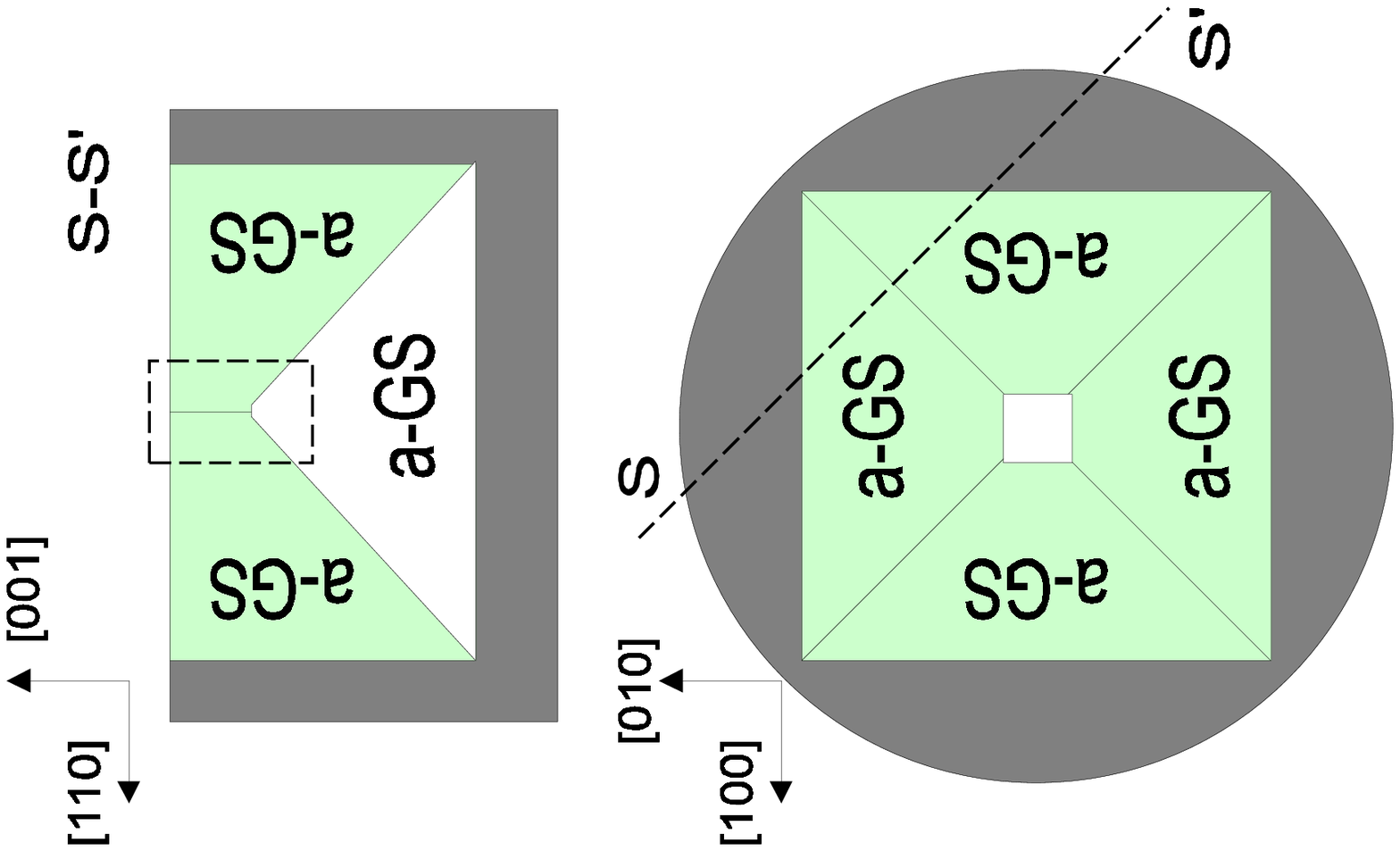} 
\hskip0.01\textwidth
\includegraphics[angle=-90,width=0.36\textwidth]{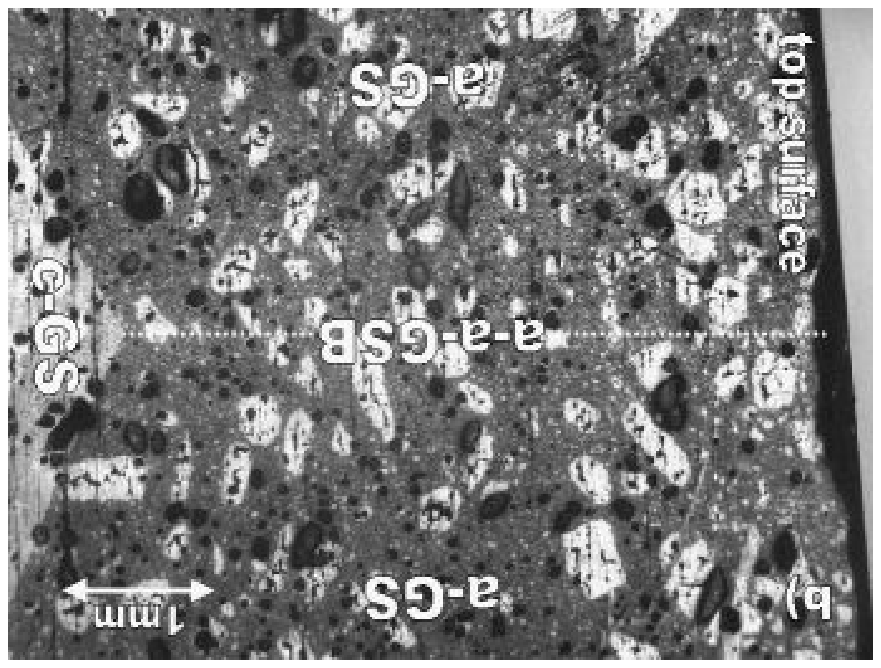} 
\end{center}
\caption{Polarized optical micrographs (a,b) of the same a-a-GSB
fragment which corresponds to the rectangle area restricted by dashed line 
in the sketch (c). The image (a) shows that the a-a-SGB is surrounded by a 
subgrain-free band, the picture (b) illustrates the 211-density difference
between the a-GSs and the c-GS (more dark regions correspond to
higher 211-density).}
\label{fig4}
\end{figure}

Having direct magnetic data about the distribution of weak links inside 
the YBCO crystal, it would be unforgivable fault to miss an opportunity of 
demonstration which structural defects do act as weak links. For this 
reason, we have cut the sample along the plane $S-S^{\,\prime}$ 
perpendicular to the a-a-GSB (see \Fref{fig4}) and studied this 
microscopic section under the polarized light. This method allows to spot 
subgrains (because a minor misalignment yields different colors and 
contrasts) as well as other imperfections of the crystalline structure. 
Following the expectations \cite{Diko}, the image (a) shows the 
subgrain-free band (with a width of approximately 1\,mm) between the 
sample surface and the c-GS apex. It confirms that weak links in the YBCO 
material are mainly related to subgrain boundaries \cite{weaklinks}.

\Fref{fig4}(b) helps to answer another important question: Why the 
geometric centre of the HTS sample, wherein the heat generation is minimal 
even at $H\gg H_{opt}$ \cite{6,Tsuchimoto}, does not correspond in 
Figures~\ref{fig3}(e)-(h) to $B_{max}$? In particular, this image shows 
that the density of the Y$_2$BaCuO$_5$ (211) inclusions, i.e. pinning 
centers responsible for the $j_c$ values, does not change across the 
a-a-GSB, but differs essentially at intersecting a-c-GSBs. Due to the 
polishing process which exposures in this place the c-GS apex (see the 
sketches presented in Figures \ref{sample} and \ref{fig4}) with a smaller 
$j_c$, this area can not carry a whole current that flows along a-a-GSBs. 
Thus, only a portion of current loops does have the X-like shape, whilst 
the others are closed around the GSB segments. That is the reason why the 
local maxima have to be exhibited by centers of these segments, but not by 
the geometric centre of the sample. Actually, the positions of the maxima 
have to move a bit closer to the crystal rims since current loops in lower 
planes, where the a-a-GSBs are edged out due to the c-GS expansion, also 
contribute to the measured flux density.

The only issue, namely, the ``dovetail''-form of isolines close to the 
crystal edges, still remains unclear. One can presume that the isoline 
forks correspond to bifurcations of the a-a-GSBs, so the forks mark the 
points where new, secondary a-GSs are spontaneously generated. However, 
this question lies outside the scope of current work, it would be the 
subject of further studies \cite{tobepublished}.

\section{Conclusions} In summary, we can conclude that the PFM does seem a 
promising method not only for a technical magnetization of HTS bulks, but 
also for their characterization: it is fast, non-destructive and, provided 
that appropriate parameters of magnetic pulse are used, very sensitive to 
minor inhomogeneities \cite{tobepublished} of the HTS material. The main 
conditions which may guarantee this sensibility are the following.
\begin{itemize}
\item The amplitude $H^*$ of magnetic pulses has to much exceed the double 
penetration field $H_p$ which linearly increases with the sample size $d$ 
and shielding properties ($\sim j_c$) of the HTS material (for example, 
$H_p$ in the \SmBCO{7-\delta} crystals is twice larger than this in the 
\YBCO ones with the same size \cite{6}). 
\item Except for the pulse amplitude, its duration $t_{max}$ and the form 
are also responsible for high velocities $V\simeq \pi\mu _0H^*/t_{max}$ of 
magnetic flux lines and, because of their interaction with viscous forces, 
considerable heat generation. Besides, short pulse duration provides 
nearly adiabatic conditions when both the heat transfer to a cooling media 
and this within the sample are negligible. 
\item The heat exchange between the sample and a cooling media has to be 
good enough to minimize the ``after-the-pulse'' smearing of the heat 
distribution. 
\end{itemize}
Since these requirements are quite easy to satisfy, this technique seems 
the most suitable to explore magnetic properties of modern, nearly perfect 
HTS crystals. Obviously, its importance should evergrow with a further 
enhancement of the HTS samples to be studied.

\ack This work was supported by the German BMBF under the project 
No.\,13N6854A3. One of the authors (A.S.) would like to thank Z.\,H.\,He 
and P.\,Diko for illuminating discussions as well as R.\,M\"uller for his 
continual encouragement throughout the present work. We are also indebted 
to M.\,Arnz and Ch.\,Schmidt for technical support.


\end{document}